\documentclass[twocolumn,floatfix,showpacs,prl]{revtex4}% 

\usepackage{epsfig}% Include figure files
\usepackage{dcolumn}% Align table columns on decimal point
\usepackage{bm}% bold math
\usepackage{amsmath}
\usepackage{graphicx}

\begin{document}

\title{Poisson transition rates from time-domain measurements with finite bandwidth\footnote{Contribution of NIST, not subject to copyright.}}

\author{O. Naaman}
\email{naaman@boulder.nist.gov}
\author{J. Aumentado}
\affiliation{National Institute of Standards and Technology, 325 Broadway, Boulder CO, 80305}%Lines break automatically or can be forced with \\

%\author{another author}

%\affiliation{another affiliation}

\date{November 4, 2005}% It is always \today, today,

\begin{abstract}
In time-domain measurements of a Poisson two-level system, the observed transition rates are always smaller than those of the actual system, a general consequence of finite measurement bandwidth in an experiment. This underestimation of the rates is significant even when the measurement and detection apparatus is ten times faster than the process under study. We derive here a quantitative form for this correction using a straightforward state-transition model that includes the detection apparatus, and provide a method for determining a system's actual transition rates from bandwidth-limited measurements. We support our results with computer simulations and experimental data from time-domain measurements of quasiparticle tunneling in a single-Cooper-pair transistor.
\end{abstract}

\maketitle

\begin{figure}[t]
\epsfxsize=3in
\epsfbox{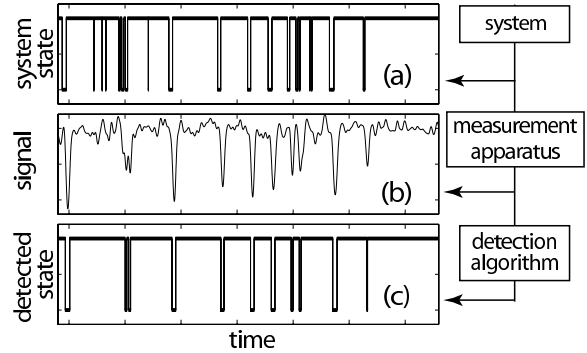}
\caption{\label{timetrace} Simulated data shown at different stages of the measurement process: (a) The true state of the system, (b) the observed telegraph signal, and (c) the detected state sequence.}
\end{figure}

We consider here a physical system switching incoherently between the states $A$ and $B$ in an alternating Poisson process, with characteristic rates $\Gamma_{A}=\tau_A^{-1}$ and $\Gamma_{B}=\tau_B^{-1}$, where $\tau_A$ and $\tau_B$ are the lifetimes of the respective states [Fig.~\ref{timetrace}(a)]. This two-state model has been used in the analysis of a wide variety of problems arising in medicine \cite{Albert98}, reliability theory \cite{QueueingBook}, network traffic \cite{Adas97}, cell physiology \cite{Neher83,Singer87}, materials science \cite{Ralls88,Stroscio04}, and condensed matter physics \cite{Lu03,Schleser04,Buehler04}, especially in relation to $1/f$ noise \cite{Machlup54}. Time-domain measurements of such systems produce random `telegraph signals' that represent the underlying transitions in the system convolved with the response of the measurement apparatus [Fig.~\ref{timetrace}(b)]. The transition rates between the states $\Gamma_A$ and $\Gamma_B$ often contain information about the underlying physical mechanism \cite{Stroscio04,Lu03,Schleser04,Buehler04}, and one of the goals of the experiment is to extract these rates from the data. To do so, one employs a detection algorithm that operates on the measured telegraph signal \cite{Lu03,Hinkley71,Yuzhelevski00,detectionBook} and, by means of a statistical test, determines the dwell times in each of the states between transitions, Fig.~\ref{timetrace}(c). The dwell times are then histogrammed to give the lifetime distribution in the two states, from which the transition rates may be determined.

A comparison of Figs.~\ref{timetrace}(a) and (c) reveals that what is observed in an experiment, \textit{i.e.}, the state of the detector, does not always reflect the true state of the system. Particularly, the bandwidth of the measurement limits the visibility of the underlying process on short time scales. In this Letter we show that the response of the measurement chain modifies the statistics of the observed process, and that the experimentally obtained transition rates always underestimate those in the underlying system. We argue that this is a general feature in finite-bandwidth time-domain measurements of stochastic two-state systems, which introduces systematic errors in estimating the actual transition rates from the data. Using a straightforward model, we calculate the lifetime distribution in the observed process, relate the experimentally observed transition rates to those in the underlying process, and compare the results of our analysis with both experimental and simulated data.

In the following we assume that the overall response of the measurement chain and the detection algorithm may be characterized by an effective detection rate, $\Gamma_{\text{det}}=1/\tau_{\text{det}}$, where $\tau_{\text{det}}$ is the mean time taken by the detector to register a transition in the measured observable after a transition in the underlying system has occurred \cite{false_alarms}. It depends on the physical bandwidth of the signal, the details of the detection algorithm, and the signal-to-noise ratio of the measurement. The statistical nature of a detector operating on a stochastic (and usually noisy) signal allows us to treat the detection itself as a stochastic process. To simplify the discussion, we will assume here that the detection process is homogeneous, so that $\Gamma_{\text{det}}$ is independent of time \cite{symmetric_rates}.

We note that the detection problem has received considerable attention in the context of patch-clamp recordings on cellular ion channels \cite{Neher83,Hawkes90,Ball88,Ball90}. These authors have tended to characterize the detector with a constant delay time, an assumption that we argue is not physical and does not agree with our experimental results shown below. Their probabilistic approach to the problem, however, allows for generalizing the analysis to processes with extended state spaces.

\begin{figure}
\epsfysize=1.3in
\epsfbox{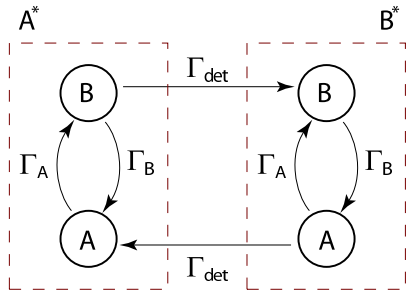}
\caption{\label{master} Schematic state diagram for the detection process.}
\end{figure}

To understand how the finite bandwidth of the measurement modifies the statistics of the observed process, let the underlying process be Poissonian, so that $t_A$ ($t_B$), the dwell time in the state $A$ ($B$), is an exponentially distributed random variable with mean $\tau_A$ ($\tau_B$). In a particular sequence, $A\rightarrow B\rightarrow A$, the excursion into $B$ might not be registered by the detector if $t_B<\tau_{\text{det}}$. Two consequences follow: 1) the lifetime distribution will have a short-time cutoff near $\tau_\text{det}$; and, more seriously, 2) the two neighboring occurrences of $A$, which were originally distinct, will be effectively `stitched' together to register a single event, longer in duration. Therefore, the \textit{observed} distribution of $t_A$ will be artificially reweighted towards longer times, with significant consequences for inferring the system parameters from the data. We will calculate this new distribution from the dynamics of the detection process, using the model shown in Fig.~\ref{master}.

In our model we separate the occurences of the states $A$ and $B$ into two manifolds, which we condition upon the state of the detector~\textendash~either $A^*$ or $B^*$, according to the state into which the last detected transition has occurred. Suppose that the detector has just entered $A^*$. We are thus on the left hand side of Fig.~\ref{master} in the state $A$. The system may then switch into state $B$, from which a transition into the $B^*$ manifold (detection) will occur at a rate $\Gamma_{\text{det}}$. While waiting to be detected, the system may cycle several times between $A$ and $B$ until a transition is eventually registered. Once in $B^*$ the cycle starts again until we return to $A^*$, and so on.     
% example of eps figure

The observed process is thus $A^*\rightleftharpoons B^*$. To find the distribution of times in $A^*$ before a transition is detected ($A^*\rightarrow B^*$), we write the rate equations that describe the evolution of the system in the left hand side of Fig.~\ref{master} (the dwell-time distribution in $B^*$ may be found by interchanging $A$ and $B$ in the following), 
\begin{equation}\label{master_eqn}
A^*\;\left\{
\begin{array}{l}
\dot{P}_A = -\Gamma_AP_A+\Gamma_BP_B,
\\ 
\dot{P}_B = -(\Gamma_B+\Gamma_{\text{det}})P_B+\Gamma_AP_A.
\end{array}\right.
%\dot{P}_{B^*}&=&\;\;\,\Gamma_{\text{det}}P_B.
\end{equation}
The experimentally accessible quantity, the dwell-time histogram, is the probability density of leaving $A^*$ in the interval $[t,t+dt]$, which is proportional to $h(t)=\Gamma_{\text{det}}P_B(t)$. Solving Eq.~(\ref{master_eqn}) with the initial conditions $P_A(0)=1$ and $P_B(0)=0$ we obtain
\begin{equation}\label{pdf}
h(t)=\frac{2}{\theta}\;\Gamma_A\Gamma_{\text{det}}\;e^{-\frac{1}{2}\lambda t}\sinh(\theta t/2),
\end{equation}
where $\lambda=\Gamma_A+\Gamma_B+\Gamma_{\text{det}}$, and $\theta=\sqrt{\lambda^2-4\Gamma_A\Gamma_{\text{det}}}$. A similar expression has been obtained in Ref.\ \cite{Ball90}. 

Eq.~(\ref{pdf}) is our main result, and has several important implications: 1) the experimentally observed process is no longer Poissonian. This should be kept in mind, for example, when the correlations \cite{Korobkova06} or the full counting statistics \cite{Gustavsson05} of the observed process are analyzed. In the short time limit, where $h(t)$ approaches zero, the ability to observe the state $B$ is bottlenecked by $\Gamma_{\text{det}}$. In the long time limit, however, Eq.~(\ref{pdf}) tends to an exponential distribution. 2) The mean dwell time in $A^*$ is 
$\langle t_{A^*}\rangle=\int{t\,h(t)dt}=\lambda/\Gamma_A\Gamma_{\text{det}}$, while the rate parameter of the long-time exponential tail of $h(t)$, which we call $\Gamma_{A^*}$, is given by
\begin{equation}\label{obs_rate_A}
\Gamma_{A^*}=\frac{\lambda}{2}\left(1-\sqrt{1-4\Gamma_A\Gamma_{\text{det}}/\lambda^2}\right).
\end{equation}
3) Researchers have traditionally interpreted both experimentally derived quantities, the lifetime $\tau_{A^*}=1/\Gamma_{A^*}$ and the mean dwell time $\langle t_{A^*}\rangle$ (which are not equal), as the Poisson lifetime $\tau_A$, but neither quantity is an accurate estimate of the true lifetime of the state $A$. For example, even when the measurement bandwidth is an order of magnitude greater than the process, say $\tau_A=\tau_B=10\,\tau_{\text{det}}$, we see that $\tau_{A^*}$ and $\langle t_{A^*}\rangle$ overestimate $\tau_A$ by 10\:\% and 20\:\% respectively. In the limit $\tau_\text{det}/\tau_A\rightarrow0$, the fractional correction to the lifetime vanishes as $\tau_\text{det}/\tau_A$. 4) Lastly, if (but only if) the detection rates are equal for both states, then Eq.~(\ref{pdf}) also implies that the equilibrium occupation probabilities in the observed and underlying processes are equal, $\bar{p}_{A^*}=\langle t_{A^*}\rangle/\left(\langle t_{A^*}\rangle+\langle t_{B^*}\rangle\right)=\bar{p}_A$. This somewhat surprising result also follows from the steady-state solution of the master equation corresponding to Fig.\ \ref{master}.

We apply the results of the above analysis, mainly Eq.\ (\ref{obs_rate_A}), to an experiment \cite{Naaman06} in which we measured the dynamics of quasiparticle (QP) tunneling in a single-Cooper-pair transistor (SCPT) \cite{Tuominen92,Joyez94,Aumentado04}. The tunneling of quasiparticles onto and off of the island of an SCPT is expected to follow a Poisson process, taking place on microsecond time scales. The transistor was operated in a regime where the island can trap a single QP \cite{Aumentado04}, so that $\Gamma_A$ and $\Gamma_B$ correspond respectively to the QP emission (un-trapping) and capture (trapping) rates. To observe this process, we used an rf reflectometry technique \cite{Schoelkopf98} with the transistor held at zero dc bias \cite{Sillanpaa04}. The reflected rf signal, whose magnitude indicates the presence or absence of a QP on the island, was measured with a spectrum analyzer. The video output of the analyzer was digitized at 500 ns intervals and recorded by a computer, where a typical record followed the evolution of the system over a span of one second. 

For this analysis, we recorded a series of time traces that differed only in the bandwidth of the intermediate-frequency (IF) filters of the spectrum analyzer. Each of these time traces was analyzed by use of both a simple hysteretic (`Schmitt trigger') detector algorithm \cite{Yuzhelevski00}, and a Page-Hinkley cumulative likelihood ratio algorithm \cite{Lu03,Hinkley71}. In both cases we set the detection thresholds so that the detector's false-alarm rate was negligible, at the expense of compromising the detection efficiency.  
\begin{figure}[t]
\epsfysize=3.5in
\epsffile{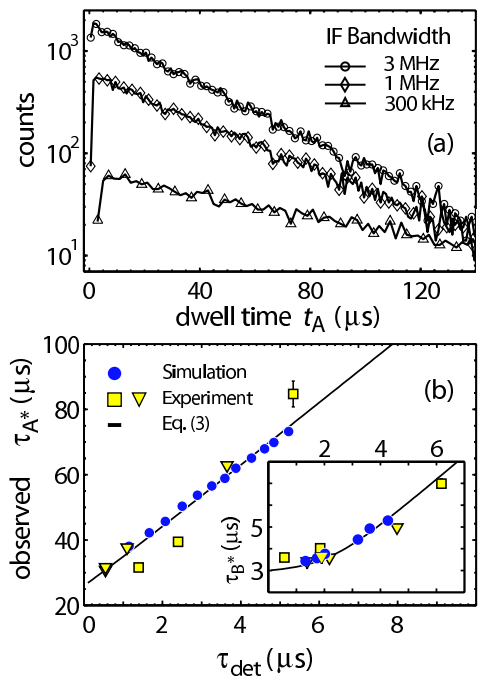} 
\caption{\label{lifetimes} (a) Histograms of QP dwell times on the island of the SCPT. The three curves correspond to different IF bandwidths-- 0.3,\,1,\,and 3\,MHz, and are offset for clarity. (b) Measured lifetimes vs.\ effective detection time for state $A$ (QP present, main panel), and state $B$ (QP absent, inset). Squares and triangle represent analysis using the Schmitt-trigger and Page-Hinkley algorithms, respectively. The solid lines were calculated using Eq.~(\ref{obs_rate_A}), and its equivalent for $\Gamma_{B^*}$, with $\tau_A=26\,\mu$s, $\tau_B=3\,\mu$s, and circles show the results of our simulations (see text).}
\end{figure}

In Figure \ref{lifetimes}(a) we show histograms of QP dwell-times on the island of the SCPT. The histograms were obtained using the Schmitt-trigger algorithm to process time traces measured with IF bandwidths of 300~kHz, 1~MHz, and 3~MHz. It is clear from the slopes of the measured distributions (on a log scale in the figure) that the observed QP tunneling rate depends on the measurement bandwidth. It appears greater when measured at a higher bandwidth, in accordance with our understanding of the detection process. Also note the sharp drop in the number of registered counts at short times, reflecting the detection bottleneck evident from Eq.~(\ref{pdf}). This short-time behavior of the dwell-time distribution might be overlooked if the data are binned too coarsely.

We obtain the QP tunneling rates, as measured with each bandwidth--algorithm combination, by fitting the lifetime distributions to an exponential. The experimental values of the lifetimes $\tau_{A^*}=1/\Gamma_{A^*}$ (QP on the island) and $\tau_{B^*}=1/\Gamma_{B^*}$ (QP off the island) are shown respectively in the main panel of Fig.~\ref{lifetimes}(b) and in the inset, plotted versus the effective detection time of the corresponding measurement. To estimate the detection time $\tau_{\text{det}}$ directly from the data, we chose to use the time at which the lifetime distribution peaks. This is only an approximation, giving an uncertainty in $\tau_{\text{det}}$ of about $\pm1~\mu$s.  

The true lifetimes are estimated by extrapolating the measured lifetimes in Fig.~\ref{lifetimes}(b) to $\tau_{\text{det}}=0$, and are approximately $\tau_A=26.0\pm0.7~\mu$s and $\tau_B=3.0\pm0.5~\mu$s. The solid curves in the figure were calculated from Eq.~(\ref{obs_rate_A}) and its equivalent for $\Gamma_{B^*}$, using the above numbers for $\tau_{A,B}$. We have also performed computer simulations, whose results are shown in Fig.~\ref{lifetimes}(b) as circles. We generated a series of $\sim10^4$ transitions between two signal levels, whose durations were taken as random variables from two exponential distributions with means equal to $\tau_{A,B}$ above. We added Gaussian noise to the signal, and applied to it a range of convolution filters with varying bandwidths. The resulting signals were then processed using our Schmitt-trigger algorithm, and analyzed as described in the preceeding paragraphs. The agreement between the simulations, experiment, and the theory, evident from the figure, indicates that our model captures the main features of finite bandwidth time-domain measurements. Characterization of the detector with a constant ``dead-time" as in Refs.\ \cite{Hawkes90,Ball88} would result in an exponential dependence of the observed lifetimes on the bandwidth of the experiment, in disagreement with our results.
\begin{figure}[t]
\epsfxsize=2.8in
\epsfbox{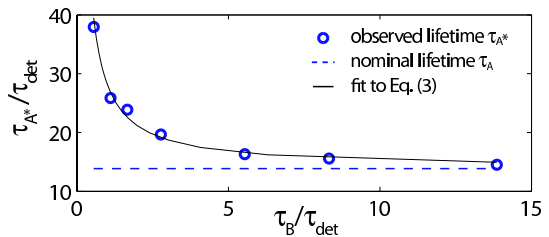}
\caption{\label{gage} Observed lifetimes $\tau_{A^*}$ from a simulated process with $\tau_A=25\,\mu$s and $\tau_B=1\;\text{to}\;25\,\mu$s. The signals were filtered with $f_c=1$\,MHz and processed with the Schmitt-trigger algorithm. From a fit (solid line) we find $\tau_{\text{det}}=1.8\,\mu$s. Because $\tau_{\text{det}}$ represents the overall response of both the filter and the algorithm, it is not surprising to find $\tau_{\text{det}}>1/f_c$.}
\end{figure}

Another set of simulations is shown in Fig.~\ref{gage}. Here we programmed a waveform generator to give an analog signal that follows a simulated Poisson process. This analog signal was used to amplitude-modulate the rf carrier in our experimental setup. We varied $\tau_B$ while holding $\tau_A$ fixed, and used a fixed filter. We plot in Fig.\ \ref{gage} the observed $\tau_{A^*}$ as circles, and the solid line is a fit to Eq.~(\ref{obs_rate_A}) with $\Gamma_{\text{det}}$ the only free parameter. A similar procedure may be used to determine the detection rate in a given experiment.
%
%\begin{figure}[t]
%\epsfxsize=2.6in
%\epsfbox{fig5.eps} 
%\caption{\label{transform} A graphical representation of the transformation $\{\tau_{A^*},\tau_{B^*}\}\rightarrow\{\tau_A,\tau_B\}$, in units of $\tau_%{\text{det}}$, using Eq.~(\ref{inverse}).}
%\end{figure}

Since only the observed process is accessible in an experiment, we proceed to find a transformation that expresses the true rates of the system in terms of the observed ones by inverting Eq.~(\ref{obs_rate_A}) and the corresponding equation for $\Gamma_{B^*}$. The inverse transformation below is given in terms of the dimensionless quantities $u=\Gamma_A/\Gamma_{\text{det}}$, $v=\Gamma_B/\Gamma_{\text{det}}$, $u^*=\Gamma_{A^*}/\Gamma_{\text{det}}$, $v^*=\Gamma_{B^*}/\Gamma_ {\text{det}}$, and is easily generalized to the case of asymmetric detection rates, 
\begin{eqnarray}\label{inverse}
u&=&\left(\frac{1-u^{*2}-v^{*2}}{1-u^*-v^*}\right)u^*-u^{*2}\nonumber
\\
v&=&\left(\frac{1-u^{*2}-v^{*2}}{1-u^*-v^*}\right)v^*-v^{*2}.
\end{eqnarray}
By use of Eq.\ (\ref{inverse}) it is straightforward to analyze the propagation of uncertainties from the measured lifetimes to the true-lifetime estimates. The lifetimes in the underlying system can alternatively be extracted from the moments of the dwell-time histograms, Eq.\ (\ref{pdf})\,: $\tau_A=\tau^{-1}_{\text{det}}\,(\langle t_{A^*}\rangle^2-\frac{1}{2}\,\langle t^2_{A^*}\rangle)$, with a similar equation for $\tau_B$.

%We use Eqs.~(\ref{inverse}) to draw lines of constant $\tau_{A^*}$ and $\tau_{B^*}$, in units of $\tau_{\text{det}}$, on the coordinate system spanned by
%the true lifetimes, as shown in Figure \ref{transform}. This figure can be used to graphically find the true lifetimes in the system under study from their %observed values. One can also get a sense from Fig.~\ref{transform} as to how uncertainties in the measured lifetimes would propagate to uncertainties in %%the true-lifetime estimates. The true lifetimes can alternatively be extracted from the moments of the dwell-time histograms, Eq.\ (\ref{pdf})\,: 
%$\tau_A=\tau^{-1}_{\text{det}}\,(\langle t_{A^*}\rangle^2-\frac{1}{2}\,\langle t^2_{A^*}\rangle)$, with a similar equation for $\tau_B$.

Although the example analyzed in this Letter is from the realm of condensed matter physics, we stress that our approach to the detection problem is applicable also in other contexts. There are a number of examples in the literature for the treatment of imperfect detectability of an underlying Markov process in medical studies \cite{Nagelkerke90}, and of Markov processes that are only partially observed, \textit{e.g.}, in reliability theory and criminology \cite{Manor98}. These authors, however, considered the detection problem as static and independent, by assigning the detector a fixed fidelity. Accounting for the dynamics of the detection process, as we have done here to address the effects of a detector's finite response time, may help identify and correct systematic errors arising in studies in those fields. A more accurate description of the detection process has been developed in the field of cellular physiology \cite{Hawkes90,Ball88,Ball90}, however, our results suggest that our stochastic model of the detector is more realistic than the ``constant dead time" model commonly used in that context. 

To conclude, we have shown that when a two-state alternating Poissonian system is measured in the time domain, the statistics of the experimentally observed process depend on the response time of the detector. Particularly, the observed process is no longer Poisson, and the transition rates that one obtains from the data always underestimate those in the underlying process. We have given analytic expressions for the relation between the experimentally observed rates and their actual values, and shown that our results are in good agreement with both experimental and simulated data. We argue that not accounting for the effects of finite measurement bandwidth will lead to results that are inaccurate at best, even when the bandwidth of the experiment is an order of magnitude greater than the underlying process. On the other hand, measurements of fast processes approaching a well characterized bandwidth limit of an experiment can still give meaningful results using the analysis presented here.

We thank R.\ L.\ Kautz, S.\ Nam, and J.\ A.\ Stroscio for valuable discussions.

%\bibliography{poisson_final_bib} % Produces the bibliography via BibTeX.

\begin{thebibliography}{29}
\bibitem[{fal()}]{false_alarms}
\bibinfo{note}{For simplicity, we neglect the detector false-alarm rate}.
\bibitem[{sym()}]{symmetric_rates}
\bibinfo{note}{We assume $\Gamma_{\text{det}}$ is equal for transitions in both
  directions. Generalizing our results for asymmetric detection rates is
  straightforward}.
\end{thebibliography}

\end{document}